**BRIEF REPORT**

# Influence of a magnetic field on the frequency of a laser stabilized to molecular iodine

Jonathan Gillot[1,2] · Joannès Barbarat[1,2] · Charles Philippe[1] · Héctor Álvarez-Martínez[1,3] · Rodolphe Letargat[1] · Ouali Acef[1]

**Abstract**
We report on the effect of a weak magnetic field applied on an iodine cell used to frequency stabilize a laser. A 1.5 µm laser is frequency tripled in order to excite the molecular transitions at 0.51 µm and frequency locked on a hyperfine line. With this frequency reference, we report short-term stability about $3 \times 10^{-14}$ $\tau^{-1/2}$, with a minimum value of $4 \times 10^{-15}$ at 200 s. The lower part of $10^{-15}$ frequency stability domain is reached, in our case, only by adding an efficient magnetic shield around the sealed quartz iodine cell. In order to quantify the Zeeman effect, we applied magnetic fields of several $\times 10^{-4}$ T on the cell containing the iodine vapour. The Zeeman effect affects the lineshape transition in such a way that we observe a modification of the laser frequency. We have measured this linear Zeeman shift at $(1062 \pm 6) \times 10^4$ Hz/T for the $a$1 hyperfine component of the R(34) 44-0 transition, near 514 nm by applying a magnetic field along the cell. Thus, in case of uncontrolled magnetic fields of an order of magnitude of $1 \times 10^{-4}$ T, the frequency stability is limited in the upper of the $10^{-14}$ domain.

## 1 Introduction

Molecular iodine represents a common interest for metrology and many experiments are dedicated to frequency stabilization applications [1, 2, 4, 11, 13]. It has been recently proposed for various applications in space, such as fundamental physics [9, 12, 14, 18, 19]. In particular, concerning the LISA project [5], the nominal solution for the reference laser in the LISA space mission is a Nd:YAG (1064.49 nm) frequency locked to an ultrastable optical cavity. The same type of Nd:YAG laser, frequency doubled and locked to a molecular iodine hyperfine line at 532.245 nm has been chosen as backup solution. Furthermore, the use of an iodine line is preferred for a transportable frequency reference for ground tests of the LISA payload. In this case, the whole setup is less sensitive to environmental parameters. Thus, SYRTE laboratory has developed a very compact (30 l), transportable and fully fibered setup based on iodine transition. This setup has been delivered to CNES, the French space agency, to perform highly precise interferometric measurements to fulfill LISA project requirements.[1] Many precautions have been taken to increase the frequency stability and maintain it at a good level for long integration times. At short time scales, the frequency instabilities are mainly limited by the product of the signal to noise ratio (SNR) and the quality factor $Q$. [23]. The signal intensity itself can be increased with the quality factor, by choosing a strong iodine transition ($Q = 2 \times 10^9$). Then, the optimization is done by the choice of the diameter of the laser beam ($\approx$ 2–3 mm), the pressure of iodine ($\geq$ 1 Pa) and the increase of the interaction length ($L_{int}$ $\geq$ 1 m). With these parameters,

✉ Jonathan Gillot
jonathan.gillot@femto-st.fr

Joannès Barbarat
joannes.barbarat@femto-st.fr

Charles Philippe
c-philippe@live.com

Héctor Álvarez-Martínez
halvarez@roa.es

Rodolphe Letargat
rodolphe.letargat@obspm.fr

Ouali Acef
ouali.acef@obspm.fr

[1] LNE-SYRTE, Observatoire de Paris, Université PSL, CNRS, Sorbonne Université, LNE, 77, avenue Denfert-Rochereau, Paris 75014, France

[2] Institut FEMTO-ST, Supmicrotech-ENSMM, UFC, CNRS, 26, rue de l'Epitaphe, Besançon 25030, Cedex, France

[3] Real Instituto y Observatorio de la Armada (ROA), San Fernando, Cadiz 11100, Spain

---

[1] To be published.

we are able to obtain stabilities at the level of a few $10^{-14}$ at short term ($\tau \simeq 1$ s) [17]. Nevertheless, the impact of the AC component of the magnetic field is rarely presented as a limitation factor in iodine-stabilized frequency references. In our case, our laboratory is situated close to the metro lines and suburban train lines and our experiment is sensitive to this magnetically noisy environment.

The Zeeman effect in molecular iodine has been largely studied [3, 22] and many experiments show the effect of a magnetic field applied on cells containing an $I_2$ vapour [8]. Since this molecule is a homonuclear diatomic one, the Zeeman effect is deemed to be very small. Given that, the control of the magnetic environment does not appear as a priority for frequency stability purpose on $I_2$ molecular transition. However, the susceptibility to the Zeeman effect depends on the sensitivity of the considered experiment. To our knowledge, the Zeeman effect in molecular $I_2$ was studied in the case of very strong magnetic fields, often between $7.8 \times 10^{-2}$ T and $5.1 \times 10^{-1}$ T [7] because the frequency stability of those experiments was not sufficient to detect the small effects of a weak magnetic field and were only sensitive to strong magnetic fields.

In this publication, we report the effect of weak magnetic fields (i.e a few gauss) on our $I_2$ frequency-stabilized system. We show the emergence of about 1 kHz frequency shift for only $1 \times 10^{-4}$ T of magnetic flux density. In other words, it contributes to the uncertainty budget with a relative frequency instability close to $10^{-12}$/G or $\approx 10^{-15}$/mG, well below some frequency stability requirements of many applications. Thus, we show that the cancellation of the AC component of the weak external magnetic fields is mandatory in order to reach the lower part of the $10^{-15}$ domain.

## 2 Methods

A RIO PLANEX laser diode emitting at 1542 nm is frequency tripled in order to excite the R(34) 44-0 $^{127}I_2$ line at 514.017 nm. We generate the third harmonic (THG) by using two lithium niobate nonlinear fibered crystals (see Fig. 1). The efficiency of this process is > 36%, so we generate 290 mW continuous wave of optical power at 514 nm from only 800 mW of infrared power at 1542 nm [16]. About 15 mW of this 514 nm light is necessary for the frequency stabilization purpose. The experimental iodine linewidth is close to 400 kHz. We use Doppler-free spectroscopy to achieve narrow hyperfine transitions in the green range, whose quality factor Q is typically greater than $2 \times 10^9$, with $Q = \nu/\delta\nu$, where $\nu$ is the center of the iodine hyperfine line, and $\delta\nu$ its full width at half maximum (FWHM). The first derivative of the $a$1 hyperfine component of the R(34) 44-0 at 514.017 nm is used to frequency stabilize the laser diode emitting at 1542.1 nm. This is done by using the well-known frequency modulation transfer technique for the iodine hyperfine line detection [20]. The frequency corrections are added to the laser diode injection current in order to frequency lock the laser. A 220 kHz frequency modulation is applied on the pump beam (3 mW) with a free space electro-optic phase modulator, while the unmodulated probe beam (0.3 mW) is split into two parts: a reference beam and an interrogating beam coupled to a 30 cm long sealed quartz cell. This cell is cooled down at −11 °C in order to reach an inner iodine pressure of 1 Pa. This is done via a cold finger which is temperature regulated at the mK level with a PID controller [10]. The pump and probe beams of $\simeq 3$ mm diameter are carefully collimated and overlapped in the cell. The polarizations of the probe and the pump are linear and orthogonal to each other. The iodine saturation signal is extracted with a balanced silicon photodiode (Fig. 2). Two acousto-optic modulators (AOMs) are used to stabilize the pump and probe optical powers [15, 21] in order to minimize light shift effect. An additional photodiode is used for the monitoring of the residual amplitude modulation (RAM) associated to the phase modulation of the pump beam, which we minimize by changing the temperature of the crystal [6]. The interaction length in the 30 cm cell is extended to 120 cm thanks to 4 successive passes.

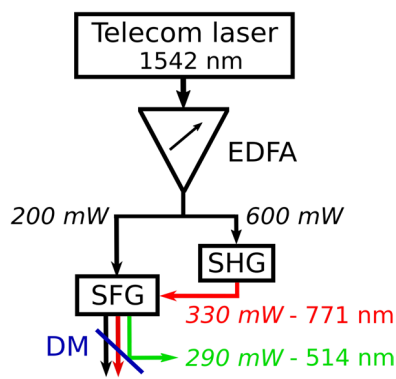

**Fig. 1** Fibered frequency tripling. After the erbium doped optical fiber amplifier (EDFA) and the splitting of the laser power, the second harmonic is generated (SHG) on the upper arm. The green light is generated as sum frequency (SFG) in a second non-linear crystal and reflected by a dichroic mirror (DM)

The frequency of our laser is then compared to that of a reference signal in order to estimate the frequency stability of our device. Our reference signal is an ultra-stable laser (USL), where a diode laser emitting at 1542 nm is frequency locked to a high finesse cavity [24]. This device is installed in another building of the Paris Observatory and we use a 200-m long fiber to bring that light to our laboratory. Then, the two signals at 1542 nm are mixed and we form a beat note having the frequency $f_{\text{bn}}$. To estimate the frequency stability, we use a K+K FXE counter with a gate time of 100

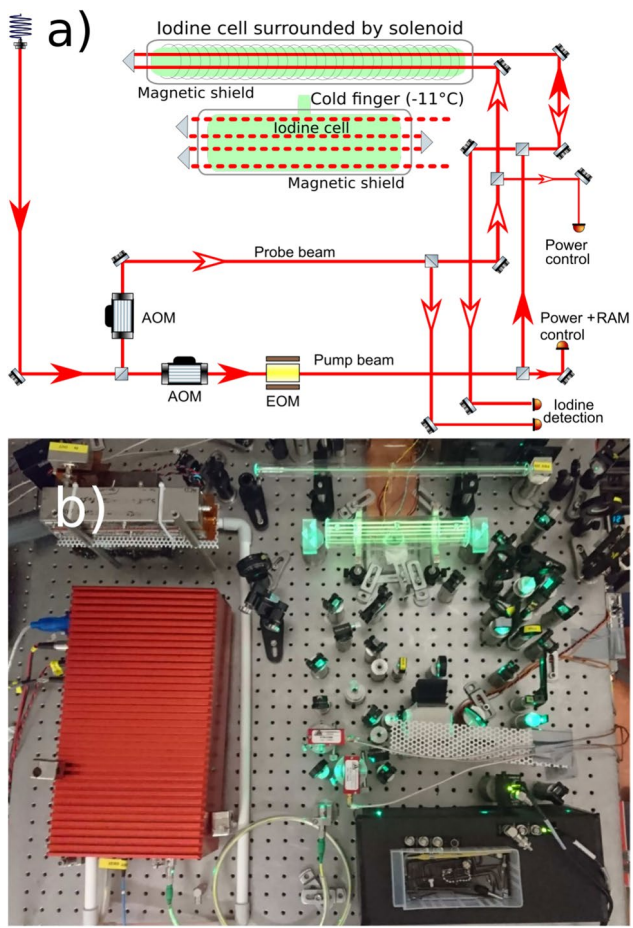

**Fig. 2** **a** Schematic representation of the iodine stabilization optical setup. *AOM* acousto-optic modulator, *EOM* electro-optic modulator. Pump and probe beams propagations are respectively indicated by plain and empty arrows. **b** Photography of the optical setup. The magnetic shields are not installed on this photography. The frequency tripling is done in the laser red box at the left

ms and measurement periods of approximatively 15 h. Typical Allan deviation is plotted on the Fig. 8. The frequency stability is close to $3 \times 10^{-14} \ \tau^{-1/2}$ with a high repeatability. A similar stability is obtained with the $a$1:P(46) 44-0 transition (514.6 nm). We have also verified that the stability is not limited by the uncompensated fiber link [15].

In order to study the Zeeman effect in molecular iodine, it is necessary to install coils to apply magnetic fields. The existence of the cold finger prevents an easy installation of the coils around the cell. For that reason, we use another non-cooled iodine cell to replace the cooled cell. This room temperature cell is made of BK7 glass and has a length of 40 cm with a 3 cm diameter. The iodine pressure is estimated to be < 2 Pa in this non-cooled cell. A typical Allan deviation obtained with this cell is presented on the Fig. 3. The effect of the magnetic field has been tested by using two independent and orthogonal coils. First, a solenoid of 1.7 mm diameter copper wire is wrapped on a BK7 glass tube. The coil includes 260 turns on the 9 cm diameter and 45 cm long BK7 tube, so the applied magnetic flux density parallel to the laser beams is given by $B_L = 7.1 \times 10^{-4}$ T/A, with values ranging from 0 to $70 \times 10^{-4}$ T. A field perpendicular to the laser beams is created by winding two coils in a quasi-rectangular shape upon the solenoid as depicted in Fig. 4. At the center of the coils, the flux density of the field behaves like $B_T = 1 \times 10^{-5}$ T/A, with values adjustable up to $3 \times 10^{-4}$ T. The solenoid has been calibrated with a compact Bartington Mag-03 3-axis magnetic sensor. The $I_2$ cell has been inserted into the glass tube so that the iodine molecules can experience a magnetic flux density in a longitudinal axis, i.e. parallel to the laser beams, and a transverse field (vertical or horizontal according to the orientation of the setup). The magnetic flux density polarity is easily reversed by changing the direction of the current.

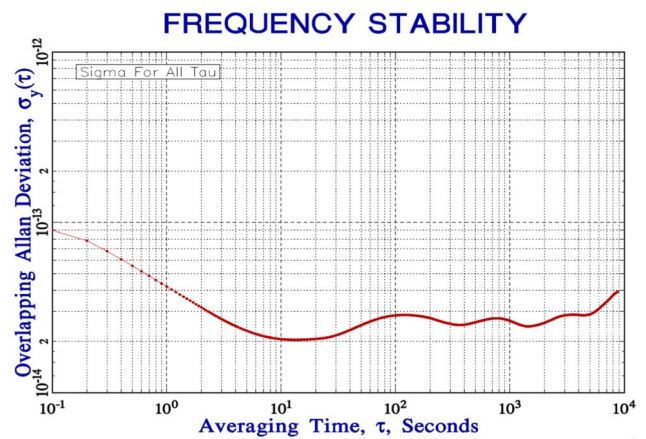

**Fig. 3** Typical Allan variance of $f_{bn}$, for an acquisition time of 0.1 s, on the room temperature iodine cell. During this measurement, the cell was not protected by a magnetic shield and was exposed to local field. This figure has been extracted from [10]

## 3 Results

The iodine molecules are also exposed to a local and variable magnetic field. This local magnetic field $B_{loc}$ is compounded of the Earth magnetic field modified by ferromagnetic objects (optical table, iron construction, etc.) and electrical currents of the public transport network. Our laboratory is situated close to the metro lines and suburban train lines, and the fluctuations of the magnetic flux density matches with the Parisian transport company schedules. The three components of $B_{loc}$ exhibit a relatively stable offset and some quick fluctuations (see Table 1). These fluctuations don't seem to own a particular frequency and stop during the night between roughly 2 am and 4 am.

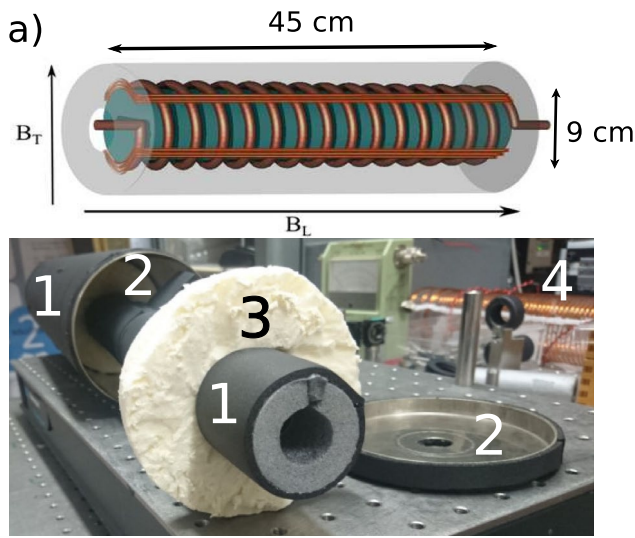

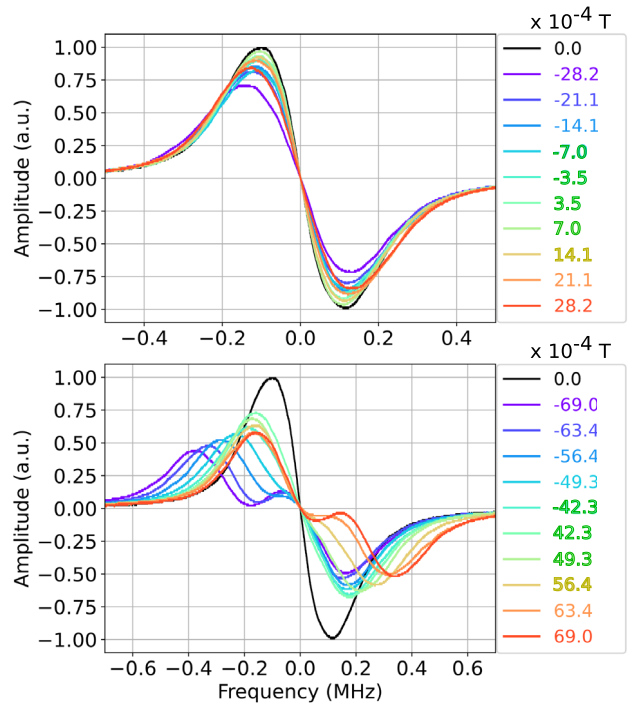

**Fig. 4** **a** A solenoid wrapped around a BK7 glass tube (in blue) creates a longitudinal $B_L$ field. Moreover, two coils are built (upon the longitudinal solenoid) in a quasi-rectangular shape to create a transverse field $B_T$. The whole setup is inserted in a magnetic shield, here in grey. The iodine cell (non visible on the scheme) is inserted into the BK7 tube. **b** Photography of the magnetic shield and of the non-magnetic mount of the cell. The dark foam is a thermal insulator (1), with one layer outside the magnetic shield and another one around the grey foam cylinder containing the cell. Two configurations are possible. The cell maintained by two white maintaining pads (3) can be introduced directly in the magnetic shield (2). Or, the iodine cell can also be introduced in the BK7 tube with copper coiled visible at the background at the right of the photography (4) and the latter introduced in the magnetic field to study the influence of a magnetic field. In that case, the iodine undergoes only the magnetic field of the coil and not the local magnetic field

**Table 1** Average values and amplitudes of the fluctuations of the three spatial components of $B_{loc}$

| Axis | DC component ($\mu$T) | Fluctuations ($\mu$T) |
| --- | --- | --- |
| North–South | − 22.8 | 0.2 |
| East–West | 0.1 | 0.2 |
| Vertical | 32.6 | 3.0 |

The North–South and East–West axis are relatively arbitrary since the magnetic probe was aligned along the weakest horizontal DC component of the magnetic field

To get rid of these local fluctuations, the setup of coils (including the $I_2$ vapour cell) is inserted into a magnetic shielding. This cylindrical shield is made of mu-metal and the measured attenuation coefficient is equal to 200. Two holes of 3 cm diameter are drilled in the two faces of the cylinder so the laser beams cross the $I_2$ vapour experiencing the applied magnetic fields $B_L$ and $B_T$. Since the cylindrical shield is oriented in the East–West axis, the residual local magnetic field $B_{loc}$ passing through the holes is on the East–West axis.

The room temperature cell has uncoated windows which induces optical feedback. In order to reduce this effect, the cell is very slightly tilted so a non-zero angle between the direction of $B_L$ and the beams appears, but this small angle is not considered in this study.

### 3.1 Effect of the magnetic field $B_L$

We have first registered the line shape of the first derivative of the $a$1: R(34) 44-0 transition. The frequency of the laser is swept while a digital scope records the saturated absorption signal. The line shape widens as $B_L$ grows as it can be seen on the upper plot in Fig. 5. For stronger fields (see lower plot in Fig. 5), the transition is split. This widening is due to the Zeeman shifting of the numerous hyperfine sub-levels of the transition. Nonetheless, our setup does not allow us to see a shift of the zero crossing of the line shape. Indeed, the natural drift of the frequency of the free running laser between two measurements can be greater than the Zeeman shifting of the zero crossing of the line shape.

In order to measure the impact of the longitudinal magnetic field $B_L$ on the frequency stability, we frequency locked

**Fig. 5** First derivative of the $a$1 hyperfine component of the R(34) 44-0 transition plotted for several values of the longitudinal flux density $B_L$ between 0 T and $29 \times 10^{-4}$ T on the upper plot. We also recorded higher values between $42 \times 10^{-4}$ T and $70 \times 10^{-4}$ T, on the lower plot. The amplitude has been scaled in regard to the measurement at 0 T

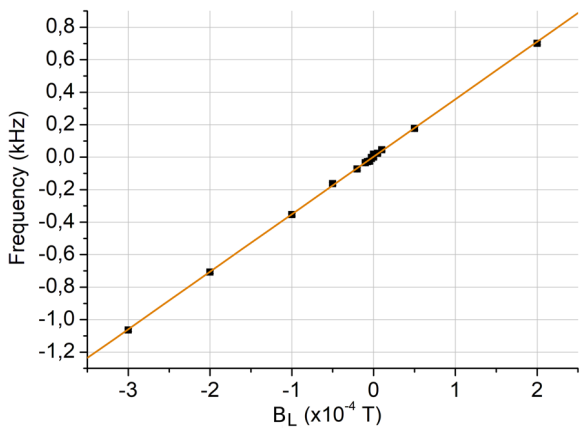

**Fig. 6** Variation of $f_{bn}$ versus $B_L$ for the a1: R(34) 44-0 transition and linearly polarized light. The sign is positive when the vector $B_L$ is antiparallel to the probe

**Table 2** Slope and offset of $f_{bn}$ versus $B_L$ for the four transitions studied with magnetic flux densities between $\pm 3 \times 10^{-4}$ T

| Transition | Slope ($\times 10^4$ Hz/T) | Offset ($\times 10^4$ Hz/T) |
| --- | --- | --- |
| a1:R(34) 44-0 | 354 ± 2 | 2 ± 2 |
| a1:R(72) 46-0 | 670 ± 1 | − 15 ± 2 |
| a1:P(90) 55-0 | 599 ± 1 | − 16 ± 2 |
| a1:R(105) 50-0 | 846 ± 2 | 14 ± 2 |

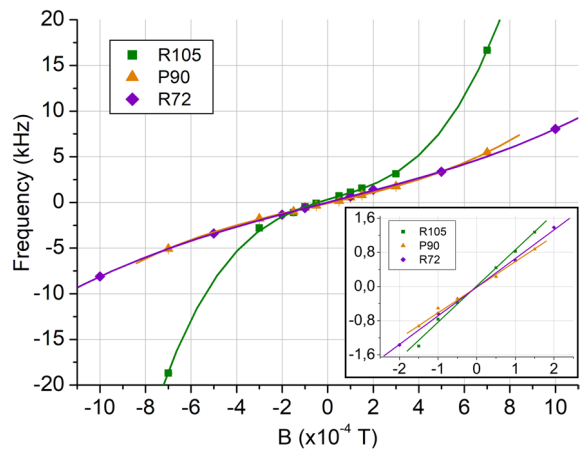

**Fig. 7** Variation of $f_{bn}$ as a function of $B_L$. The sign of $B_L$ is a convention. The flux density of $B_L$ has been varied from $-1 \times 10^{-3}$ T to $+1 \times 10^{-3}$ T. On the insert, we zoomed on the values between $-2.5 \times 10^{-4}$ T et $+2.5 \times 10^{-4}$ T in which the applied magnetic field is weaker and the frequency dependence is linear

the laser and we applied several current values between 0 and 3 A with a 10 mA uncertainty in the solenoid. We synchronously register the beat note frequency $f_{bn}$. The value of the current is changed every 5 min, and for each value of $B_L$, $f_{bn}$ remains stable. The frequency stability of the laser during each time interval between the changes of the electric current is $\simeq 2.8 \times 10^{-14} \tau^{-1/2}$. $f_{bn}$ exhibits a remarkable change each time we vary the current, in correlation with the magnetic flux density of $B_L$. An independent beat note with another reference signal allow us to keep track of the frequency of the USL, so we can easily discriminate a frequency fluctuation coming from the cavity or a change of our laser frequency.

Moreover, we can assuredly reject the effect of a cavity drift. The ultra-stable cavity is subject to a maximum drift of 100 mHz/s, due to slight thermal fluctuations resulting in non linear frequency behaviour of the cavity. We regularly apply a zero magnetic field so we can estimate the cavity drift by fitting the frequency values for $B_L$=0. Knowing its value, we can readily remove it. Anyway, the drift is very small compared to the fluctuations of $f_{bn}$ due to the changes of the applied magnetic field. We focused in the $\pm 3 \times 10^{-4}$ T region, where the magnetic field is comparable to the magnetic field in the laboratory (see Fig. 6). As the dependence seems to be proportional, we made a linear fit of the experimental data. The line has a slope of $(354 \pm 2) \times 10^4$ Hz/T with an offset very close to zero $(2 \pm 2)$ Hz. The zero of frequency has been arbitrary chosen for $f_{bn}$ with $B_L = 0$ T. The remarkable linearity of the effect is pointed out by $R^2$, the square of the sample correlation coefficient, which is close to 0.99. The beat note was generated at 1542nm, but the molecular iodine transition is at 514 nm, thereby the shift of the latter is $(1062 \pm 6) \times 10^4$ Hz/T. In terms of frequency stability, we get $\Delta \nu / \nu = (1.82 \pm 0.01) \times 10^{-8}$ per tesla of longitudinal field. Taking into account the fluctuations of the local magnetic field due to human activities shown in the Table 1, we can estimate their impact on the frequency stability. For an amplitude of the fluctuations of $2 \times 10^{-7}$ T along the horizontal axis, the frequency stability is going to be limited close to $\Delta \nu / \nu = (3.64 \pm 0.02) \times 10^{-15}$.

Furthermore, we extended our study at other hyperfine iodine lines with different $J$ and $v$ quantum numbers, $a1$:R(72) 46-0, $a1$:P(90) 55-0 and $a1$:R(105) 50-0 as shown in Fig. 7 and Table 2. We observe that the Zeeman frequency shift depends on the value of $J$ and $v$, and seems to be increasing with $J$ in strong field regime. This point is not surprising because Zeeman sub-levels with greater quantum number are subject to a higher Zeeman shifting [8]. The iodine molecule offers transitions that are more or less sensitive to the external magnetic field. However, the choice of the useful transition is dictated by its SNR and its linewidth with the aim of frequency stabilization. For that reason, the rest of this study is done with the R(34) 44-0.

For higher magnetic flux densities, we observe a non-linear Zeeman effect beyond $\pm 3 \times 10^{-4}$ T, more or less important depending on the molecular transition (see Fig. 7).

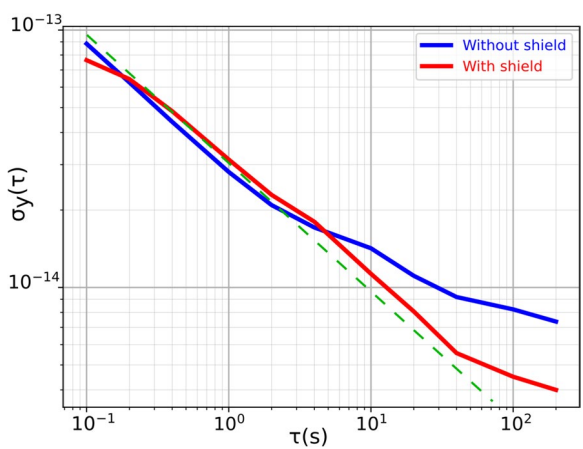

**Fig. 8** Typical Allan variance of $f_{bn}$. The blue curve corresponds to the cooled iodine cell without magnetic shield and exposed to local field. At long term, the Allan deviation exhibits a plateau in the upper $10^{-15}$ range. The red curve corresponds to the cooled iodine cell when inserted in the described magnetic shield. The long term Allan variance is in the low $10^{-15}$ range. The green dashed curve is the $3 \times 10^{-14}\ \tau^{-1/2}$ slope

### 3.2 Effect of the magnetic field $B_T$

The maximum transverse flux density achievable with a current of 10 A and a 5 turns coil is of $1.11 \times 10^{-4}$ T and it was not sufficient to quantify its effect on the line shape. However, we detected the impact of $B_T$ on the frequency stability. If we assume a linear dependence with the flux density of the magnetic field, like for the longitudinal case, the frequency beat note $f_{bn}$ exhibits a drift of $(60 \pm 25) \times 10^4$ Hz/T at 1542 nm, i.e. $(180 \pm 75) \times 10^4$ Hz/T at 514 nm. This corresponds to $\Delta\nu/\nu = (3.09 \pm 1.29) \times 10^{-9}$ per tesla of transverse field, supposing a linear dependence. Considering the fluctuations of $B_{loc}$ given in the Table 1, we can estimate their impact on the frequency stability. The spatial component of $B_{loc}$ subject to the biggest fluctuations is the vertical one. For an amplitude of the fluctuations of $3 \times 10^{-6}$ T, the frequency stability is going to be limited close to $\Delta\nu/\nu = (9.26 \pm 3.86) \times 10^{-15}$.

### 3.3 Impact on frequency stability

A simple way to check these results is to compare the frequency stability with and without the magnetic shield, for $B_T$=0 and $B_L$=0. As expected with a magnetic shielding, the frequency stability is better at long term, i.e. after 5 s. According to the dependencies of $(1062 \pm 6) \times 10^4$ Hz/T and $(180 \pm 75) \times 10^4$ Hz/T found for respective longitudinal and transverse fields, the Allan variance is in the lower $10^{-15}$ range. At this level, other effects like the uncompensated fiber link between the USL and our laser are no longer negligible (Fig. 8).

## 4 Conclusion

By applying a weak magnetic field to the molecular iodine cell, we measured a frequency shift of $(1062 \pm 6) \times 10^4$ Hz/T at 514 nm for the longitudinal field $B_L$. In regards to the transverse field $B_T$, the frequency shift measured as preliminary is $(180 \pm 75) \times 10^4$ Hz/T at 514 nm.

We have shown that the Zeeman effect due to a weak magnetic field is far from being negligible. Indeed, for several applications, like LISA the required frequency stability is 30 Hz/$\sqrt{\text{Hz}}$. Thereby, weak magnetic field fluctuations of an order of magnitude of $1 \times 10^{-5}$ T can deteriorate the frequency stability below the requirements.

Our experiment highlighted a broadening of the line shape of the transition and also a shift of the zero crossing of the line shape. Thus, unstable environmental magnetic fields can cause instabilities in iodine frequency stabilized lasers. In a more fundamental point of view, it appears clearly that frequency-stabilized experiments for metrology are useful to quantify small effects, such as Zeeman effect on molecular species considered as sparely sensitive to magnetic fields.

Considering the measurements reported in this paper, a special twin magnetic shielding ensuring an attenuation factor $\geq 1000$ has been designed to reduce the contribution of the magnetic field to residual frequency instabilities of the stabilized laser at the $10^{-16}$ level.

**Acknowledgements** We are indebted to J. Hrabina from ISI Lab (CZ) for providing uncooled borosilicate iodine cell used in this work. We would like to thank F. Du Burck from LPL laboratory (Université Paris-Nord) for many discussions and advices on iodine spectroscopy. We thank also our colleague E. De Clercq for his help and discussions during the experiment. We are indebted to the electronics staff and the mechanical workshop for their technical support and highlighting contributions. The authors, and especially J. Gillot and J. Barbarat, would like to thank the national network for time and frequency LabEx FIRST-TF for giving us the opportunity to work on this project and for the funding, as well as the DGA (ANR STABI2-011-001) and the CNES for the funding of the thesis of C. Philippe. The authors also thank the anonymous referees for their insightful suggestions.

The data that support the findings of this study are available from the corresponding author upon reasonable request. The authors declare no competing interests.

**Author contributions** Jonathan Gillot wrote the main manuscript text and Joannès Barbarat prepared the Figs. 2, 5 and 8. Ouali Acef decided the organization of the manuscript. Hector Alvarez-Martinez and Rodolphe Letargat provided some data, reviewed the manuscript and brought substantial modifications. All authors reviewed the manuscript.

**Data availability** The data that support the findings of this study are available from the corresponding author upon reasonable request.

**Declarations**

**Conflict of interest** The authors declare no Conflict of interest.

**Publisher's Note** Springer Nature remains neutral with regard to jurisdictional claims in published maps and institutional affiliations.